\newcommand{\extended}[1]{}  

\documentclass{llncs}

\usepackage{amssymb,amsmath}
\usepackage{tikz}
\usetikzlibrary{arrows,automata,shapes,snakes,backgrounds,petri,mindmap,fit,positioning}
\usepackage{xspace}
\usepackage{color}
\usepackage[utf8x]{inputenc}
\usepackage[T1]{fontenc}
\usepackage{graphicx}
\usepackage{subfigure}
\usepackage{multirow}
\usepackage{multicol}
\usepackage{algorithm}
\usepackage{algorithmic}
\usepackage{threeparttable}
\usepackage{latexsym}
\usetikzlibrary{shapes,arrows,automata}
\usetikzlibrary{calc}

\usepackage{wojtek15logics,wojtek15other}
{\begin{figure*}[#1]
\centering
\begin{minipage}{#2}
\begin{algorithm}[H]}%
{\end{algorithm}
\end{minipage}
\end{figure*}}

\newcommand{\castleddefeated}{\mathit{castle3defeated}}
\newcommand{\alldefeated}{\mathit{allDefeated}}
\newcommand{\tianji}{\mathit{TJ}}
\newcommand{\tianjiwins}{\mathit{TJWins}}
\newcommand{\tianjiwonraces}{\mathit{TJWonLess2}}
\newcommand{\AEMC}{\lan{AE\ensuremath{\mu}C}}
\newcommand{\valuation}{\ensuremath{\mathcal{V}}}
\newcommand{\Vars}{\ensuremath{\mathcal{V}\textit{ar}}}
\newcommand{\denot}[3]{\ensuremath{[\![#1]\!]^{#2}_{#3}}}
\newcommand{\Mintersect}{M_1}

\title{Approximating Strategic Abilities under Imperfect Information: a Naive Approach}
\author{
  Wojciech Jamroga
  \and
  Michal Knapik
  \and
  Damian Kurpiewski
}
\institute{
  Institute of Computer Science, \\ Polish Academy of Sciences, Poland \\
  \email{\{w.jamroga,michal.knapik,damian.kurpiewski\}@ipipan.waw.pl}
}

\begin{document}
\maketitle

\begin{abstract}
Alternating-time temporal logic (\ATL) allows to specify requirements on abilities that different agents should (or should not) possess in a multi-agent system. However, model checking \ATL specifications in realistic systems is computationally hard. In particular, if the agents have imperfect information about the global state of the system, the complexity ranges from \Deltwo to undecidable, depending on the syntactic and semantic details. The problem is also hard in practice, as evidenced by several recent attempts to tackle it.
On the other hand, model checking of alternating epistemic mu-calculus can have a distinctly lower computational complexity.
In this work, we look at the idea of approximating the former problem by the verification of its ``naive'' translations to the latter. In other words, we look at what happens when one uses the (incorrect) fixpoint algorithm to verify formulae of \ATL with imperfect information.
\end{abstract}

\section{Introduction}

There is a growing number of works that study syntactic and semantic variants of \emph{strategic logics}, in particular the alternating-time temporal logic \ATL. Conceptually, the most interesting strand builds upon reasoning about temporal patterns and outcomes of strategic play, limited by information available to the agents.
The contributions are mainly theoretical, and include results concerning the conceptual soundness of a given semantics of ability~\cite{Schobbens04ATL,Jamroga03FAMAS,Agotnes04atel,Jamroga04ATEL}, meta-logical properties~\cite{Bulling14comparing-jaamas}, and the complexity of model checking~\cite{Schobbens04ATL,Jamroga06atlir-eumas,Jamroga06mis-tr}.
However, there is very little research on the actual \emph{use} of the logics, in particular on practical algorithms for reasoning and/or verification.

This is somewhat easy to understand, since model checking of \ATL variants with imperfect information has been proved \Deltwo- to \Pspace-complete for agents playing positional (a.k.a.~memoryless) strategies~\cite{Schobbens04ATL,Jamroga06atlir-eumas} and undecidable for agents with perfect recall of the past~\cite{Dima11undecidable}.
Moreover, the imperfect information semantics of \ATL does not admit fixpoint equivalences~\cite{Bulling14comparing-jaamas}, which makes incremental synthesis of strategies impossible, or at least difficult to achieve.
Some practical attempts at tackling the problem started to emerge only recently~\cite{Pilecki14synthesis,Busard14improving,Huang14symbolic-epist}.
Up until now, experimental results confirm that the initial intuition was right: model checking strategic modalities for imperfect information is hard, and dealing with it requires innovative algorithms and verification techniques.

One idea that has not been properly explored is that of alternating-time epistemic mu-calculus (\AEMC)~\cite{Bulling11mu-ijcai}. Since fixpoint equivalences do not hold under imperfect information, it follows that standard fixpoint translations of \ATL modalities lead to a different interpretation of strategic ability. In fact, it can be argued that they capture existence of \emph{recomputable} winning strategies. However, what especially interests us in the context of model checking is that they can make model checking computationally cheaper. Verification of \AEMC is in general between \NP and \Deltwo, but the scope of backtracking is much smaller than for \ATL with imperfect information, as it includes only the actions starting from a given indistinguishability class rather than all the actions in the model. Moreover, for coalitions of up to 2 agents the model checking problem is in \Ptime~\cite{Bulling11mu-ijcai}.

The question that we ask in this paper is: Is \AEMC an attractive alternative for verification of strategic abilities under imperfect information?
To this end, we will look at the naive \AEMC approximations of formulae of \ATL[ir] (i.e., \ATL with imperfect information and imperfect recall), and investigate:
\begin{enumerate}
\item Whether model checking of the \AEMC approximations performs significantly faster than for the original \ATL[ir] formulae;
\item Whether the \AEMC counterparts are indeed semantic approximations of the \ATL[ir] specifications, or they encapsulate a completely different concept of ability.
    As fixpoint equivalences are not valid for \ATL[ir], we know that the naive fixpoint translation is in general incorrect. However, one can possibly ask: how often?
\end{enumerate}

We take on an empirical approach.
More precisely, we consider two classes of benchmark models and formulae, one based on the Tian Ji scenario~\cite{Lomuscio15mcmas,Busard14improving} and the other being the Castles benchmark from~\cite{Pilecki14synthesis}.
Then, for a formula $\varphi$, we compare the output and performance of the \ATL[ir] model checking of $\varphi$ with the \AEMC model checking of $aemc(\varphi)$, i.e., with the straightforward (and generally incorrect) fixpoints approximation of $\varphi$. The work reported here is very preliminary, but it already allows to draw some conclusions, and decide on the most promising lines for future research.

\section{What Agents Can Achieve under Imperfect Information}

In this section we provide a brief overview of the relevant variants of \ATL, and the corresponding complexity results for model checking. We refer the interested reader to~\cite{Alur02ATL,Schobbens04ATL,Bulling11mu-ijcai} for details.

\subsection{Models}

The semantics for \ATL is defined over a variant of transition systems where transitions are labeled with combinations of actions, one per agent. An \emph{imperfect information concurrent game structure} (ICGS)~\cite{Alur02ATL,Schobbens04ATL} is given by
$\model = \tuple{\Agt, \States, \Pi,
\pi, Act, d, o,\{\sim_a \mid a\in \Agt\}}$
which includes a nonempty finite set of all agents
$\Agt = \set{1,\dots,k}$, a nonempty set of states $\States$, a set
of atomic propositions $\Pi$ and their valuation
$\pi:\Props\rightarrow \powerset{\States}$, and a nonempty finite
set of (atomic) actions $\Actions$. Function $d : \Agt \times
\States \rightarrow \powerset{Act}$ defines nonempty sets of actions
available to agents at each state, and $o$ is a (deterministic)
transition function that assigns the outcome state $q' =
o(q,\alpha_1,\dots,\alpha_k)$ to state $q$ and a tuple of actions
$\langle\alpha_1, \dots, \alpha_k\rangle$ for $\alpha_i \in d(i,q)$
and $1\leq i\leq k$, that can be executed by $\Agt$ in $q$. We  write $d_a(q)$ instead of $d(a,q)$. Each $\sim_a\subseteq
\States\times\States$ is an equivalence relation satisfying  $d_a(q)=d_a(q')$ for
$q\sim_a q'$.
Note that perfect information can be modeled by assuming each $\sim_a$ to be the minimal reflexive relation.

\begin{figure}[t]\centering
\begin{tikzpicture}[>=latex,scale=1.4]
\input{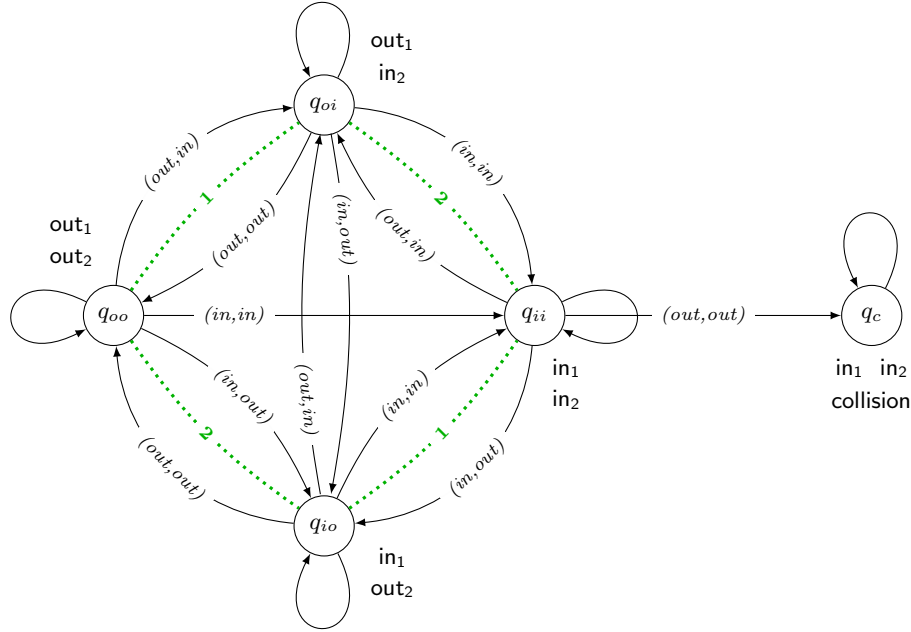}
\end{tikzpicture}
%
\caption{Autonomous vehicles at the intersection: model $\Mintersect$}
\label{fig:intersection}
\end{figure}

\begin{example}[Intersection with limited visibility]\label{ex:intersection-imperf}
Consider an intersection with $k$ autonomous vehicles around it. Each vehicle is modeled as a separate agent, whose local state is characterized by either the proposition \prop{out_i} (when the vehicle is outside the intersection) or \prop{in_i} (when the vehicle is inside it). The available actions are: $in$ (``drive in'' or ``stay in'', depending on the current state) and $out$ (``drive out'' or ``stay out'').
Transitions update the state accordingly, except for one case: when both agents are in and decide to leave at the same time, a collision occurs (\prop{collision}).
Furthermore, let us assume that no agent sees the location of the other vehicle.

Figure~\ref{fig:intersection} presents a pointed ICGS modeling the scenario for $k=2$. The combinations of actions that are not displayed in the graph do not change the state of the system.
The indistinguishability relations are depicted by dotted lines.
\qed
\end{example}

A \emph{strategy} of agent $a$ is a conditional plan that specifies
what $a$ is going to do in each situation.
Here, we only refer to \emph{memoryless uniform strategies} (\emph{\ir strategies} in short), defined as functions $s_a : \States\rightarrow Act$ such that $s_a(q)\in d_a(q)$ for all $q$, and $q\sim_a q'$ implies $s_a(q)=s_a(q')$.
A \emph{collective strategy} $s_A$ is a tuple of \ir strategies, one per agent from $A$.
%
A \emph{path} $\lambda=q_0q_1q_2\dots$ is an infinite sequence of
states such that there is a transition between each $q_i,q_{i+1}$.
We use $\lambda[i]$ to denote the $i$th position on path $\lambda$
(starting from $i=0$) and $\lambda[i,\infty]$ to denote the subpath
of $\lambda$ starting from $i$.
%
Function $out_\model(q,s_A)$ returns the set of all paths that can result from the execution of strategy $s_A$ from state $q$ in model $\model$, defined formally as follows:
\begin{description}
\item[$out(q,s_A) =$]  $\{ \lambda=q_0,q_1,q_2\ldots \mid
      q_0=q$ and for each $i=0,1,\ldots$ there exists
      $\tuple{\alpha^{i}_{a_1},\ldots,\alpha^{i}_{a_k}}$ such that
      $\alpha^{i}_{a} \in d_a(q_{i})$ for every $a\in \Agt$,
      and $\alpha^{i}_{a} = s_A[a](q_{i})$ for every $a\in A$,
      and $q_{i+1} = o(q_{i},\alpha^{i}_{a_1},\ldots,\alpha^{i}_{a_k}) \}$.
\end{description}
Moreover, we define $out^{ir}_\model(q,s_A) = \bigcup_{a\in A}\bigcup_{q\sim_a q'}\out_\model(q',s_A)$.
We will omit the subscript $\model$ if it is clear from the context.

\subsection{Alternating Time Temporal Logic}

Let $\Agt$ be the set of agents and $\Props$ be the set of atomic propositions. The language of \ATL is given by the following grammar:\\
$$\varphi ::= p  \mid  \lnot \varphi  \mid  \varphi\land\varphi  \mid
  \coop{A} \Next\varphi  \mid  \coop{A} \Always\varphi  \mid
  \coop{A} \varphi\Until\varphi$$
where $A\subseteq \Agt$ and $\prop{p}\in\Props$. Additionally, we define ``sometime in the future'' as $\Sometm\varphi \equiv \top\Until\varphi$.
The semantics of \ATL[ir] is defined by the following clauses:
\begin{description2}
\item[]$M,q \models p$\quad iff $q \in \pi(p)$ \qquad (where $p \in \Pi$);

\item[]$M,q \models \neg\varphi$\quad iff $M, q \not\models \varphi$;

\item[]$M,q \models \varphi \land \psi$\quad iff $M, q \models \varphi$
    and $M, q \models \psi$;

\item[]$M,q \models \coop{A}\Next\varphi$\quad iff there is a collective strategy $s_A$
  such that, for each $\lambda\in out^{ir}(q,s_A)$,  $M, \lambda[1] \models \varphi$;

\item[]$M,q \models \coop{A}\Always \varphi$\quad iff there is a collective strategy $s_A$
  such that, for each $\lambda\in out^{ir}(q,s_A)$ and  every $i\ge 0$, $M,\lambda[i] \models \varphi$;

\item[]$M,q \models \coop{A}\varphi\Until\psi$\quad iff there is $s_A$
  such that, for each $\lambda\in out^{ir}(q,s_A)$, there is $i\ge0$ for which
  $M,\lambda[i]\models\psi$, and $M,\lambda[j]\models\varphi$ for each $0\le j<i$.
\end{description2}
Informally, $\model,q \models \coop{A}\gamma$ iff there exists a strategy for $A$ such that $\gamma$ holds on all the paths that the agents in $A$ consider as possible executions of the strategy.

\begin{example}[Intersection with limited visibility, ctd.]
Take model $\Mintersect$ from Example~\ref{ex:intersection-imperf}.
Now, we have e.g.~that
$\Mintersect,q_{oo} \models \coop{1}\Always\neg\prop{collision}$ (it suffices that agent 1 executes action ``$out$'' regardless of anything).
On the other hand,
$\Mintersect,q_{oo} \models \neg\coop{1,2}\Sometm\prop{collision}$ (the agents do not know how to make sure that a collision will happen, even if they want to).
We leave it up to the interested reader to check the latter.
\qed
\end{example}

\subsection{Verification of Strategic Abilities}

The model checking problem asks, given a model $M$, a state $q$ in it, and a logical formula $\varphi$, whether $\varphi$ holds in $M,q$.
\ATL verification is known to be tractable for perfect information models, but intractable for imperfect information.

\begin{proposition}[\cite{Schobbens04ATL,Jamroga08mcheckcloser}]
Model checking of \ATL[ir] is \Deltwo-complete in the number of states and transitions in the model, and the length of the formula.
\end{proposition}

\subsection{Alternating Epistemic Mu-Calculus}

Alternating epistemic $\mu$-calculus (\AEMC) replaces the temporal-strategic operators $\coop{A}\Always,\coop{A}\Until$ with the \emph{least fixpoint operator} $\mu$~\cite{Bulling11mu-ijcai}: \\
$$\varphi ::= p  \mid  Z \mid  \lnot \varphi  \mid  \varphi\land\varphi  \mid
  \coop{A} \Next\varphi \mid \mu Z. \varphi$$ 
where $Z\in\Vars$ is a second order variable ranging over sets of states.
The \emph{greatest fixed point operator} $\nu$ can be defined as dual to $\mu$.
We consider only the \emph{alternation-free} fragment of $\AEMC$, cf.~\cite{Alur02ATL,Bulling11mu-ijcai} for the exact definition.

A \emph{valuation of $\Vars$} is a mapping $\valuation:\Vars\rightarrow\powerset{\States}$.
Given a set $Q\subseteq\States$ of states we define $\valuation[Z:=Q]$ as the update of $\valuation$ that assigns $Q$ to $Z$. The semantics of \AEMC is given by the denotation function $\denot{\varphi}{M}{\valuation}$ which defines the set of states in $M$ that satisfy $\varphi$ in the following way:\
$\denot{\prop{p}}{M}{\valuation}=\pi(\prop{p})$,\  $\denot{Z}{M}{\valuation}=\valuation(Z)$,\  $\denot{\neg\varphi}{M}{\valuation}=\States\setminus \denot{\varphi}{M}{\valuation}$,\ $\denot{\varphi\wedge\psi}{M}{\valuation}=\denot{\varphi}{M}{\valuation}\cap \denot{\psi}{M}{\valuation}$,\ $\denot{\coop{A}\Next\varphi}{M}{\valuation}=\{q\mid \exists \alpha_A\in d_A(q)\forall \alpha_{\Agt\backslash A}\in d_{\Agt\backslash A}: o(q,(\alpha_A,\alpha_{\Agt\backslash A}))\in\denot{\varphi}{M}{\valuation}\}$,\ and $\denot{\mu{}Z.\varphi}{M}{\valuation}=\bigcap\{Q\subseteq \States\mid \denot{\varphi}{M}{{\valuation[Z:=Q]}}\subseteq Q\}$.
%
%
Moreover, $\model,q\models\varphi$ iff $q\in\denot{\varphi}{\model}{\valuation}$ for all valuations $\valuation$.

\begin{proposition}[\cite{Bulling11mu-ijcai}]
Model checking of \AEMC is \Ptime-complete if all the coalitions in $\varphi$ consist of at most 2 agents.
For abilities of coalitions with $3$ or more agents, it is between \NP and \Deltwo in the size of the largest abstraction class of relations $\sim_1,\dots,\sim_k$.
\end{proposition}

Thus, verification of \AEMC is potentially more attractive than \ATL[ir].
A natural idea is to use the naive translation of \ATL[ir] to \AEMC, defined as follows:
\begin{eqnarray*}
aemc(p) &=& p \\
aemc(\neg\varphi) &=& \neg aemc(\varphi) \\
aemc(\varphi\land\psi) &=& aemc(\varphi)\land aemc(\psi) \\
aemc(\coop{A}\Next\varphi) &=& \coop{A}\Next aemc(\varphi) \\
aemc(\coop{A}\Always\varphi) &=& \nu Z.(aemc(\varphi) \land \coop{A}\Next Z) \\
aemc(\coop{A}\Sometm\varphi) &=& \mu Z.(aemc(\varphi) \lor \coop{A}\Next Z) \\
aemc(\coop{A}\varphi\Until\psi) &=& \mu Z.(aemc(\psi) \lor aemc(\varphi)\land\coop{A}\Next Z).
\end{eqnarray*}
We will look at how it works in the subsequent sections.

Note that, in a way, $aemc(\coop{A}\Sometm\varphi)$ expresses a \emph{persistent} ability to achieve $\varphi$.
Likewise, $aemc(\coop{A}\Always\varphi$ expresses a {persistent} ability to maintain $\varphi$.
This is because $aemc(\coop{A}\Sometm\varphi)$ produces a strategy for $A$ such that $aemc(\coop{A}\Sometm\varphi)$ will also hold for every state reachable by the strategy. For $aemc(\coop{A}\Always\varphi)$ the situation is analogous.

\section{Fixpoint Approximation of Strategic Ability: Performance}

%
%
%

In this section, we empirically compare the performance of model checking \ATL[ir] specifications vs. their naive \AEMC approximations.

The experimental results on the \AEMC side have been obtained by running a straightforward implementation of the fixpoint model checking algorithm, implemented in Python 3.
The tests have been conducted on a MacBook with an Intel Core i5 CPU with dynamic clock speed of 1.4 GHz,
4 GB of RAM (one module DDR3, 1600 MHz buz clock), and OS X 10.10.5 Yosemite.

The performance of \ATL[ir] model checking for the Castles model is cited after~\cite{Pilecki16synthesis}, and was obtained on a notebook with an Intel Core i7-3630QM CPU with dynamic clock speed of $2.4$ GHz up to $3.4$ GHz, and 8 GB of RAM (two modules DDR3 PC3-12800, 800 MHz bus clock, effective data rate 1600 MT/s, in dual-channel configuration). Two model checkers were used: SMC and MCMAS.
The experiments with SMC were conducted on Windows 7 OS, the experiments with MCMAS on Linux Ubuntu 12.04.2.
Thus, the \ATL[ir] model checking for Castles was performed on a significantly better computing equipment than the \AEMC verification.
We also note that SMC uses several reduction techniques to restrict the search space, and MCMAS operates on compact symbolic representations of models, based on BDD's. In contrast, our model checking of \AEMC was done with a straightforward implementation of the standard explicit state algorithm with no optimizations at all.

The performance of \ATL[ir] model checking for the TianJi model is cited after~\cite{Busard14improving}, and was obtained with an experimental model checker implemented with PyNuSMV, a Python framework for prototyping and experimenting with BDD-based model-checking algorithms based
on NuSMV~\cite{Busard13PyNuSMV}.\footnote{
  It should be mentioned that the results for TianJi in~\cite{Busard14improving} were obtained for a slightly different semantics of \ATL[ir], employing additional fairness constraints. }
Besides compact symbolic representations of states and transitions, the model checker features multiple optimization techniques. The authors do not describe the computing configuration that was used for their experiments.

The timeout in all cases is defined as 120 minutes.

\begin{figure}[t]
\centering
\begin{tabular}{|c||c|c|c||c||c|}
\hline
 & \multicolumn{3}{|c||}{\AEMC($\psi_1'$)} & \ATL[ir]/SMC($\psi_1$) & \ATL[ir]/MCMAS($\psi_1$) \\ \hline\hline
{\bf Configuration} & {\bf time} & {\bf \#sat} & {\bf \#iter} & {\bf time} & {\bf time} \\ \hline
4 (1,1,1)       & 0.011         & 128   & 1       & timeout & 72 \\ \hline
5 (1,1,2)       & 0.024      & 256   & 1       & timeout & timeout \\ \hline
6 (2,1,2)       & 0.386     & 512   & 1       & timeout & timeout \\ \hline
7 (2,2,2)       & 9.231    & 1024  & 1       & timeout & timeout \\ \hline
8 (3,2,2)       & 4352.891 & 5504  & 2        & timeout & timeout \\ \hline
\end{tabular}
\caption{Model checking performance for Castles, formula $\psi_1'$ vs.~$\psi_1$}
\label{fig:castlesres1}
\end{figure}

\begin{figure}[t]
\centering
\begin{tabular}{|c||c|c|c||c||c|}
\hline
 & \multicolumn{3}{|c||}{\AEMC($\psi_2'$)} & \ATL[ir]/SMC($\psi_2$) & \ATL[ir]/MCMAS($\psi_2$) \\ \hline\hline
{\bf Configuration} & {\bf time} & {\bf \#sat} & {\bf \#iter} & {\bf time} & {\bf time} \\ \hline
4 (1,1,1)             & 0.004          & 8           & 1               & timeout & 78     \\ \hline
5 (1,1,2)             & 0.015      & 16          & 1                & timeout & error    \\ \hline
6 (2,1,2)             & 0.050      & 32          & 1                & ? & ?    \\ \hline
7 (2,2,2)             & 0.225     & 64          & 1                 & ? & ?   \\ \hline
8 (3,2,2)             & 1.202    & 128         & 1                  & ? & ?  \\ \hline
\end{tabular}
\caption{Model checking performance for Castles, formula $\psi_2'$ vs.~$\psi_2$. ``?'' indicates lack of data about the performance of the given model checker on the given instance. }
\label{fig:castlesres2}
\end{figure}

\subsection{Benchmark 1: Castles}

The \emph{Castles} model have been proposed in~\cite{Pilecki14synthesis}.
The model consists of one agent called \emph{Environment} that keeps track of the health points of three castles, plus a number of agents called \emph{Workers} each of whom works for the benefit of a castle.
Health points (HP, ranging from $0$ to $3$) represent the current condition of the castle; $0$ HP means that the castle is \emph{defeated}.
Workers can execute the following actions:
attack a castle they do not work for,
defend the castle they do work for, or
do nothing.
Doing nothing is the only available action to a Worker of a defeated castle.
No agent can defend its castle twice in a row, it must wait one step before being able to defend again.
A castle gets damaged if the number of attackers is greater than the number of defenders, and the damage is equal to the difference.
In the initial state, all the castles have 3 HP and every Worker can engage in defending its castle.
The indistinguishability relations for Workers are defined as follows. Every Worker knows if it can currently engage in defending its castle, and can observe for each castle if it is defeated or not.
The model is parameterized by the number of agents and the allocation of Workers. For example, an instance with 1 worker assigned to the first castle, 3 workers assigned to the second
and 4 to the third castle will be denoted by \textit{9 (1,3,4)}.

\subsubsection{Formulae}

We considered the following formulas for Castles:
\begin{align*}
\psi_1 & \equiv \coop{c12}\Sometm\castleddefeated\\
\psi_2 & \equiv \coop{w12}\Sometm\alldefeated
\end{align*}
The first formula says that the workers working for castles 1 and 2 have a collective strategy to defeat castle 3, no matter what other agents do.
Similarly, the second formula says that workers number 1 and 2 have a collective strategy to ensure the defeat of all castles.
After the naive translation to \AEMC we obtain:
\begin{align*}
\psi_1' & \equiv \mu Z. (\castleddefeated \lor \coop{c12}\Next Z)\\
\psi_2' & \equiv \mu Z. (\alldefeated \lor \coop{w12}\Next Z)
\end{align*}

\subsubsection{Experimental Results}

The results for Castles are presented in Figures~\ref{fig:castlesres1} and ~\ref{fig:castlesres2}.
The tables present results for a sequence of models of various sizes.
The column headers are interpreted as follows:
\begin{itemize}
\item time: model checking time (in seconds),
\item \#sat: number of states in which the formula is satisfied,
\item \#iter: number of iterations until reaching fixpoint. Note: we only have data about \#sat and \#iter for the experiments with \AEMC model checking.
\end{itemize}

\begin{figure}[t]
\centering
\begin{tabular}{|c||c|c|c||c|}
\hline
 & \multicolumn{3}{|c||}{\AEMC($\phi_1'$)} & \ATL[ir]($\phi_1$) \\ \hline\hline
{\bf Horses} & {\bf time} & {\bf \#sat} & {\bf \#iter} & {\bf time} \\ \hline
3            & 0.0002 & 11 & 2  &   2.603     \\ \hline
4            & 0.001 & 18 & 2  &    8.205    \\ \hline
5            & 0.014 & 153 & 3  &   30.885     \\ \hline
6            & 0.024 & 300 & 3  &   99.931     \\ \hline
7            & 0.753 & 2258 & 4   & 586.126      \\ \hline
8            & 6.204 & 4900 & 4   &   ?    \\ \hline
\end{tabular}
\caption{Model checking performance for TianJi, formula $\phi_1'$ vs.~$\phi_1$}
\label{fig:tianji1}
\end{figure}

\begin{figure}[t]
\centering
\begin{tabular}{|c||c|c|c||c|}
\hline
 & \multicolumn{3}{|c||}{\AEMC($\phi_2'$)} & \ATL[ir]($\phi_2$) \\ \hline\hline
{\bf Horses} & {\bf time} & {\bf \#sat} & {\bf \#iter} & {\bf time} \\ \hline
3            & 0.0002 & 0 & 2  &   $\approx 2.5$     \\ \hline
4            & 0.001 & 0 & 2  &    $\approx 10$    \\ \hline
5            & 0.003 & 0 & 2  &    $\approx 650$    \\ \hline
6            & 0.004 & 0 & 2  &    ?    \\ \hline
7            & 0.016 & 0 & 2   &   ?    \\ \hline
8            & 0.075 & 0 & 2   &   ?    \\ \hline
\end{tabular}
\caption{Model checking performance for TianJi, formula $\phi_2'$ vs.~$\phi_2$}
\label{fig:tianji2}
\end{figure}

\subsection{Benchmark 2: TianJi}

The second series of experiments has been conducted for the TianJi variant from~\cite{Busard14improving}.
The model consists of two agents: Tian Ji and the king.
Each agent has $n$ horses numbered $1,\dots,n$.
In the game, Tian Ji and the king send their horses one by one against each other.
Horse $i$ of Tian Ji wins the race with king's horse $j$ iff $i>j$.
At each stage, the agents know the current score and their own remaining horses, but not those of the opponent.
Moreover, the decisions at each round are made simultaneously, so one does not know which horse is currently sent by the other player.
The player whose horses won most races wins the game.


We considered the following formulas for TianJi:
\begin{align*}
\phi_1 & := \coop{\tianji}\Sometm\tianjiwins\\
\phi_2 & := \coop{\tianji}\Always\coop{\tianji}\Next\tianjiwonraces
\end{align*}
$\tianjiwins$ holds when the game is done and TianJi has won more races than the king.
Similarly, $\tianjiwonraces$ is satisfied when TianJi has won at most 1 race up to the current point.
After the naive translation we obtain:
\begin{align*}
\phi_1' & := \mu Z. (\tianjiwins \lor \coop{\tianji}\Next Z)\\
\phi_2' & := \nu Z. (\coop{\tianji}\Next\tianjiwonraces \land \coop{\tianji}\Next Z)
\end{align*}
The results of experiments are presented in Figures~\ref{fig:tianji1} and~\ref{fig:tianji2}.

\subsection{Discussion}

The experimental results show that, for the instances of model checking that we have tested, verification of strategic abilities in alternating epistemic $\mu$-calculus offers a dramatic speedup over model checking ``standard'' \ATL with imperfect information. So, from the computational point of view, \AEMC is definitely more attractive than \ATL[ir].
The speedup occurs despite the fact that the \AEMC model checking was done by a straightforward implementation of the standard explicit state algorithm, while the results for \ATL[ir] were obtained by model checkers that use multiple optimization techniques and, in most cases, also compact symbolic representation of states and transitions in the model.

We also note that \AEMC model checking was distinctly faster than that of \ATL[ir] even when they were both (theoretically) \NP-complete, i.e., for coalitions larger than 2 agents (cf.~Figure~\ref{fig:castlesres1}). 

\begin{figure}[t]
\centering
\begin{tabular}{|c||c||c|}
\hline
{\bf Configuration} & \multicolumn{1}{|c||}{\AEMC($\psi_1'$)} & \multicolumn{1}{|c|}{\ATL[ir]($\psi_1$)} \\ \hline\hline
4 (1,1,1)       & false  & true \\ \hline
5 (1,1,2)       & false  & true \\ \hline
6 (2,1,2)       & false  & true \\ \hline
6 (3,1,1)       & true   & true \\ \hline
7 (2,2,2)       & false  & true \\ \hline
8 (3,2,2)       & true   & true \\ \hline
\end{tabular}
\caption{Model checking output for Castles, formula $\psi_1'$ vs.~$\psi_1$}
\label{fig:castlesres1-outp}
\end{figure}

\begin{figure}[t]
\centering
\begin{tabular}{|c||c||c|}
\hline
{\bf Configuration} & \multicolumn{1}{|c||}{\AEMC($\psi_2'$)} & \multicolumn{1}{|c|}{\ATL[ir]($\psi_2$)} \\ \hline\hline
4 (1,1,1)             & false    & false     \\ \hline
5 (1,1,2)             & false     & false    \\ \hline
6 (2,1,2)             & false     & false    \\ \hline
6 (3,1,1)             & false     & false    \\ \hline
7 (2,2,2)             & false      & false   \\ \hline
8 (3,2,2)             & false       & false  \\ \hline
\end{tabular}
\caption{Model checking output for Castles, formula $\psi_2'$ vs.~$\psi_2$}
\label{fig:castlesres2-outp}
\end{figure}

\section{Fixpoint Approximation of Strategic Ability: Semantics}

In this section, we empirically compare the output of model checking \ATL[ir] with that of model checking \AEMC. That is, we compare the truth values of \ATL[ir] specifications in the benchmark models, versus the truth values of their naive fixpoint approximations.
The output of \AEMC model checking has been produced by our model checking algorithm. 
The truth values according to the \ATL[ir] semantics were determined by hand.

\subsection{Benchmarks 1 \& 2: Castles and TianJi}

The results for Castles are presented in Figures~\ref{fig:castlesres1-outp} and ~\ref{fig:castlesres2-outp}.
The tables present the truth values of a given formula in the initial state of the benchmark model of a given size.
Similarly, the output of model checking for TianJi is presented in Figures~\ref{fig:tianji1-outp} and~\ref{fig:tianji2-outp}.

\begin{figure}[t]
\centering
\begin{tabular}{|c||c||c|}
\hline
{\bf Horses} & \multicolumn{1}{|c||}{\AEMC($\phi_1'$)} & \multicolumn{1}{|c|}{\ATL[ir]($\phi_1$)} \\ \hline\hline
3            & false  &  false   \\ \hline
4            & false  &  false   \\ \hline
5            & false  &  false   \\ \hline
6            & false  &  false   \\ \hline
7            & false   & false   \\ \hline
8            & false   & false   \\ \hline
\end{tabular}
\caption{Model checking output for TianJi, formula $\phi_1'$ vs.~$\phi_1$}
\label{fig:tianji1-outp}
\end{figure}

\begin{figure}[t]
\centering
\begin{tabular}{|c||c||c|}
\hline
{\bf Horses} & \multicolumn{1}{|c||}{\AEMC($\phi_2'$)} & \multicolumn{1}{|c|}{\ATL[ir]($\phi_2$)} \\ \hline\hline
3            & false  &   false   \\ \hline
4            & false  &   false   \\ \hline
5            & false  &   false   \\ \hline
6            & false  &   false   \\ \hline
7            & false   &  false   \\ \hline
8            & false   &  false   \\ \hline
\end{tabular}
\caption{Model checking output for TianJi, formula $\phi_2'$ vs.~$\phi_2$}
\label{fig:tianji2-outp}
\end{figure}

\subsection{Benchmark 3: Modified TianJi}

The experiments with Castles have shown that the \AEMC approximations capture a much more restrictive notion of ability than the original \ATL[ir] specifications.
In this view, the results in Figures~\ref{fig:tianji1-outp} and~\ref{fig:tianji2-outp} are not very informative.
Both semantics have produced the same truth values, but was it because they are indeed so close? Or rather because the \ATL[ir] semantics incidentally produced ``false,'' i.e., the truth value that the fixpoint semantics seems to favor?

To answer this question, we have prepared and executed an additional run of experiments based on a modification of the TianJi story. In ``Modified TianJi,'' general TianJi always sees the horse selected by the king before sending his own horse to the next race. The modification significantly increases the strategic abilities of the general.
The results are presented in Figure~\ref{fig:modtianji-outp}.

\begin{figure}[t]
\centering
\begin{tabular}{|c||c||c|}
\hline
{\bf Horses} & \multicolumn{1}{|c||}{\AEMC($\phi_1'$)} & \multicolumn{1}{|c|}{\ATL[ir]($\phi_1$)} \\ \hline\hline
3            & false  &    true   \\ \hline
4            & false  &    true   \\ \hline
5            & false  &     true   \\ \hline
6            & false  &    true   \\ \hline
7            & false   &    true   \\ \hline
8            & false   &    true   \\ \hline
\end{tabular}
\quad
\begin{tabular}{|c||c||c|}
\hline
{\bf Horses} & \multicolumn{1}{|c||}{\AEMC($\phi_2'$)} & \multicolumn{1}{|c|}{\ATL[ir]($\phi_2$)} \\ \hline\hline
3            & false  &    true   \\ \hline
4            & false  &    true   \\ \hline
5            & false  &    true   \\ \hline
6            & false  &     true   \\ \hline
7            & false   &    true   \\ \hline
8            & false   &    true   \\ \hline
\end{tabular}
\caption{Model checking output for Modified TianJi}
\label{fig:modtianji-outp}
\end{figure}

\subsection{Discussion}

The experiments show that using \AEMC to approximate model checking of \ATL[ir] in a straightforward way does not work. There is no correlation between the truth of the \ATL[ir] formulae that we have tested, and their naive \AEMC translations. In fact, the latter ones did not hold in an overwhelming majority of models that we looked at. This is due to the fact that \emph{persistent} or \emph{recomputable} strategic ability is a much stronger property than being able to come up with a winning strategy only in the initial state of the game. 

This suggests two possible ways of further study.
One is to identify subclasses of concurrent game structures where recomputable strategies can be obtained. Models of agents with perfect recall seem a natural candidate in this respect.
The other is to suitably weaken the \AEMC translations so that they capture also existence of (some) non-recomputable strategies.
We leave exploration of both paths for future research.

\section{Conclusions}

In this paper, we have looked at verification of strategic abilities for agents with imperfect information. The aim was to investigate whether straightforward fixpoint approximations provide an interesting alternative to formulae of \ATL[ir], for which model checking is known to be theoretically and practically hard.
The answer is both yes and no. 
On one hand, our experimental results show that verification of ``fixpoint abilities,'' specified in alternating epistemic $\mu$-calculus, offers a dramatic speedup over model checking of \ATL[ir]. On the other hand, there is no correlation between satisfaction of the \ATL[ir] formulae that we have tested, and their naive \AEMC translations. 
Thus, we conclude that \AEMC \emph{is} an attractive alternative to \ATL[ir] from the computational point of view, but it does \emph{not} approximate model checking of \ATL[ir] in a straightforward way.
Rather, it is underpinned by a distinctly different notion of ability, based on existence of \emph{persistent} or \emph{recomputable} strategies.

In the future, we plan to identify subclasses of concurrent game structures where such recomputable strategies can be obtained. 
We will also investigate how to weaken the fixpoint translations so that they capture also existence of some non-recomputable strategies.

\bibliographystyle{plain}
\bibliography{wojtek,wojtek-own}

\begin{thebibliography}{10}

\bibitem{Agotnes04atel}
T.~{\AA}gotnes.
\newblock A note on syntactic characterization of incomplete information in
  {ATEL}.
\newblock In {\em Procedings of Workshop on Knowledge and Games}, pages 34--42,
  2004.

\bibitem{Alur02ATL}
R.~Alur, T.~A. Henzinger, and O.~Kupferman.
\newblock {A}lternating-time {T}emporal {L}ogic.
\newblock {\em Journal of the ACM}, 49:672--713, 2002.

\bibitem{Bulling11mu-ijcai}
N.~Bulling and W.~Jamroga.
\newblock Alternating epistemic mu-calculus.
\newblock In {\em Proceedings of {IJCAI-11}}, pages 109--114, 2011.

\bibitem{Bulling14comparing-jaamas}
N.~Bulling and W.~Jamroga.
\newblock Comparing variants of strategic ability: How uncertainty and memory
  influence general properties of games.
\newblock {\em Journal of Autonomous Agents and Multi-Agent Systems},
  28(3):474--518, 2014.

\bibitem{Busard14improving}
S.~Busard, C.~Pecheur, H.~Qu, and F.~Raimondi.
\newblock Improving the model checking of strategies under partial
  observability and fairness constraints.
\newblock In {\em Formal Methods and Software Engineering}, volume 8829 of {\em
  Lecture Notes in Computer Science}, pages 27--42. Springer, 2014.

\bibitem{Busard13PyNuSMV}
Simon Busard and Charles Pecheur.
\newblock Pynusmv: Nusmv as a python library.
\newblock In {\em Proceedings of {NASA Formal Methods}}, pages 453--458, 2013.

\bibitem{Dima11undecidable}
C.~Dima and F.L. Tiplea.
\newblock Model-checking {ATL} under imperfect information and perfect recall
  semantics is undecidable.
\newblock {\em CoRR}, abs/1102.4225, 2011.

\bibitem{Huang14symbolic-epist}
X.~Huang and R.~van~der Meyden.
\newblock Symbolic model checking epistemic strategy logic.
\newblock In {\em Proceedings of {AAAI}}, pages 1426--1432, 2014.

\bibitem{Jamroga03FAMAS}
W.~Jamroga.
\newblock Some remarks on alternating temporal epistemic logic.
\newblock In B.~Dunin-Keplicz and R.~Verbrugge, editors, {\em Proceedings of
  Formal Approaches to Multi-Agent Systems (FAMAS 2003)}, pages 133--140, 2003.

\bibitem{Jamroga06mis-tr}
W.~Jamroga and T.~{\AA}gotnes.
\newblock Modular interpreted systems: A preliminary report.
\newblock Technical Report IfI-06-15, Clausthal University of Technology, 2006.

\bibitem{Jamroga06atlir-eumas}
W.~Jamroga and J.~Dix.
\newblock Model checking {ATL$_{ir}$} is indeed {$\Delta_2^P$}-complete.
\newblock In {\em Proceedings of EUMAS'06}, volume 223 of {\em {CEUR} Workshop
  Proceedings}. CEUR-WS.org, 2006.

\bibitem{Jamroga08mcheckcloser}
W.~Jamroga and J.~Dix.
\newblock Model checking abilities of agents: A closer look.
\newblock {\em Theory of Computing Systems}, 42(3):366--410, 2008.

\bibitem{Jamroga04ATEL}
W.~Jamroga and W.~van~der Hoek.
\newblock Agents that know how to play.
\newblock {\em Fundamenta Informaticae}, 63(2--3):185--219, 2004.

\bibitem{Lomuscio15mcmas}
A.~Lomuscio, H.~Qu, and F.~Raimondi.
\newblock {MCMAS}: An open-source model checker for the verification of
  multi-agent systems.
\newblock {\em International Journal on Software Tools for Technology
  Transfer}, 2015.
\newblock To appear.

\bibitem{Pilecki14synthesis}
J.~Pilecki, M.A. Bednarczyk, and W.~Jamroga.
\newblock Synthesis and verification of uniform strategies for multi-agent
  systems.
\newblock In {\em Proceedings of {CLIMA XV}}, volume 8624 of {\em Lecture Notes
  in Computer Science}, pages 166--182. Springer, 2014.

\bibitem{Pilecki16synthesis}
J.~Pilecki, M.A. Bednarczyk, and W.~Jamroga.
\newblock Synthesis and verification of uniform strategies for multi-agent
  systems.
\newblock 2016.
\newblock Journal version, under submission.

\bibitem{Schobbens04ATL}
P.~Y. Schobbens.
\newblock Alternating-time logic with imperfect recall.
\newblock {\em Electronic Notes in Theoretical Computer Science}, 85(2):82--93,
  2004.

\end{thebibliography}

\end{document}